\newcommand{\ignore}[1]{}
\documentstyle[12pt,epsf]{article}                                                  
        
\begin{document}                                                              
\begin{titlepage}                                                              

\begin{flushright}                                                            
\hfil                 BROWN-HET-984\\  
\hfil                 BU-CCS-950602\\                           
\hfil                 May 1995 \\                             
\end{flushright}                                             
\begin{center}                                                          
{\Large\bf Monte  Carlo Study of the  Yukawa Coupled \\
\vskip .5cm
Two Spin Ising Model}


\vspace{1.5cm}            

\large{
Richard Brower$^{a}$, Kostas Orginos$^{b}$, Yue Shen$^{a}$ and
Chung-I Tan$^{b}$\\
}
\end{center}                                    
$^{a}$Department of Physics, Boston University, Boston, MA 02215\\
$^{b}$Department of Physics, Brown University, Providence, RI\\
\vspace{1.9cm}                                                    

\abstract{ We consider a particular  4 state  spin system 
composed of two Ising spins (~$s_x, \; \sigma_x$~) with independent hopping
parameters $\kappa_1, \kappa_2$, coupled by a bilinear Yukawa term, $y s_x
\sigma_x$. The Yukawa term is solely responsible for breaking the global $ Z_2
\times Z_2$ symmetry down to $Z_2$. This model is intended as an illustration
of general coupled Higgs system where scalars can arise both as composite and
elementary excitations. For the Ising example in 2d, we give convincing
numerical evidence of the universality of the two spin system with the one
spin Ising model, by Monte Carlo simulations and finite scaling analysis . We
also show that as we approach the phase transition, universality arises by a
separation of low mass spin waves from an extra set of spin waves with
an energy gap that diverges as the correlations length diverges.}

\ignore{
We study a spin model composed of two Ising models with hopping
parameters $\kappa_1, \kappa_2$, respectively which are coupled
linearly to each other with coupling $y$. We find that the continuum
limit of this model is in the same universality class as the 2d Ising
model. In particular, when $ \beta \equiv \kappa_1 = \kappa_2 = y/2$,
the model reduces to a 3 dimensional Ising model with two layers in
the third dimension. The critical point is at $\beta_c = 0.2759(7)$
and all critical exponents are consistent with a 2d Ising model.  
}
\vfill                                                                    

\end{titlepage}

\section{Introduction}

It often happens in quantum field theory that exactly the same physical system
can be described by very different actions. For example, the critical behavior
of a two dimensional paramagnetic film can be described either by a Bosonic 2d
Ising model, or equivalently by a 2d free Fermion field \cite{Kogut}.  Another
example is the exact equivalence of Sine-Gordan theory and the massive
Thirring model. Kinks in the elementary scalar field theory become elementary
Fermions while the scalars are composite Fermion anti-Fermion bound
states\cite{Coleman}.  

A similar equivalence exists for the Goldstone modes of chiral symmetry
breaking which can be described either in terms of an elementary scalar field
theory or as composite scalar modes in a Fermion-antiFermion channel. For
example it has been shown that the Higgs-Yukawa model and the generalized
Nambu-Jona-Lasinio Model \cite{UCSD} are identical. Note that in the last
example, the number of degrees of freedom appears to be very different at the
cut-off scale, yet the low energy physics turns out to be exactly the
same. All of these are manifestations of the the well known universality
property: different microscopic Lagrangians can lead to the same continuum
theory (critical behavior) as long as the fixed point structure and the low
mass excitations of the two actions are the same.

Here we wish to consider a different but closely related phenomenon where the
same excitation is represented twice in a single Lagrangian by two different
fields, yet in the scaling limit the theory is universal to either spin system
taken separately. Such a redundant description has recently been advocated as
an improved lattice formulation of GCD in the so called chirally extended
action (XQCD)\cite{BST},\cite{BOT}.  Here an elementary scalar field is
coupled to the Wilson lattice QCD action.  It was conjectured that XQCD is in
the same universality class as the Wilson lattice QCD action and its critical
property describes the continuum QCD physics. An obvious problem for XQCD is
the extra degrees of freedom introduced by the scalar fields on the lattice.
Will the extra degrees of freedom decouple in the continuum limit? This
question can not be answered easily without nonperturbative calculations.

An important feature of this suggestion is the existence of a global (Higgs)
symmetry G in each field theory so that the combined theory has a $G \times G$
symmetry that is broken down to $G$ by a Yukawa coupling.  In these more
complicated examples, it is not obvious when they remain in the same
universality class as the original theory or if numerical methods can give
convincing evidence. We were able to prove that an extended Nambu-Jona Lasino
model and a two spin coupled O(N) models are equivalent in large
N~\cite{BST,BOT}.  At large N, the extra degrees of freedom become
decoupled in the continuum limit and the low energy spectrum is exactly the
same as if no extra degrees of freedom were introduced at the cut-off
scale. Our purpose in this article is to begin the use of numerical methods to
supplement analytical methods.  To this end we consider a simple model of
redundant coupled Ising spins to see if Monte Carlo methods can settle the
question of universality and expose the mechanism by which it is achieved in
the scaling limit.

\section{Yukawa Coupled Spin Ising Model}

For the our  two spin Ising system, we consider the action,
\begin{equation}
S = -\kappa_1\sum_{x,\mu} \sigma_{x + \mu}\sigma_x
-\kappa_2\sum_{x,\mu} s_{x + \mu}s_x
- y \sum_x \sigma_x s_x~,
\label{eq:model}
\end{equation}
\noindent where $\sigma_x = \pm 1$ and $s_x = \pm 1$.
The sum over $x = (n_1, n_2)$ is performed on a two dimensional lattice, $n_i
= 1,2,\cdots, L$, with nearest neighbor couplings, $\mu = (1,0), \; (0,1)$.  A
very general class of four spins models could be considered. Our particular
choice we call a ``Yukawa'' coupling borrowing terminology from field theories
for spontaneous breaking of continuous symmetry. When the Yukawa coupling y is
set to zero, this model (\ref{eq:model}) reduces to two independent 2d Ising
models with a global $Z_2 \times Z_2$ symmetry.  But when $y$ is different
from zero, only a single $Z_2$ symmetry is survives.  This is a special
example of 4 state spin models, which to our knowledge has no analytic
solution.  It is no longer self-dual.

At the cut-off scale there are two fields, $\sigma_x$ and $s_x$, in
the coupled Ising model in Eq. (\ref{eq:model}). However, as we will
show in the following sections, only one eigenvalue of the mass matrix
remains in the spectrum in the continuum limit. The other mass
eigenvalue stays at the cut-off scale. We believe that the light mass
corresponds to the massive particle of a 2d Ising model.  We will
further show that the coupled Ising model follows the scaling laws of
a 2d Ising model. Therefore, our model in Eq. (\ref{eq:model}) is in
the same universality class of a 2d Ising model.

Admittedly in this case the universality of our 4 state spin model to the
Ising model is not surprising.  The Yukawa coupled two spin model is in fact
nothing but an Ising model written on 3-d lattice with only two layers in the
third axis and three independent couplings: $\kappa_1$ and $\kappa_2$ in each
x-y layer and $y$ along the z axis. Since this model can be viewed as a $d = 2
+ \epsilon$ model with vanishing $\epsilon$ in the thermodynamic limit and
since other anisotropic Ising models are known to be in the isotropic Ising
universality class, universality with a single Ising spin system appears
likely . When we tune to the critical surface and the correlation length $\xi$
becomes much larger than $2$, the system should behave like a single layer 2d
Ising model. Still it is important for our larger purposes to see if numerical
methods can give convincing evidence and that analysis of the spectrum can
illuminate the mechanism for approaching the single spin system in the scaling
limit. Finally we should note that since the totally decoupled system ($y =
0$) is clearly not a single Ising system, so one expects that by carefully tuning
the Yukawa term a new interacting two spin fixed point must also exist which
is not universal to a single spin system.  This fine tuning limit is
what is often required to obtain a true 2 Higgs system in the continuum.

\section{Phase Diagram}

\begin{figure}
\epsfxsize=11.0cm
\epsfysize=9.0cm
$$
\epsfbox{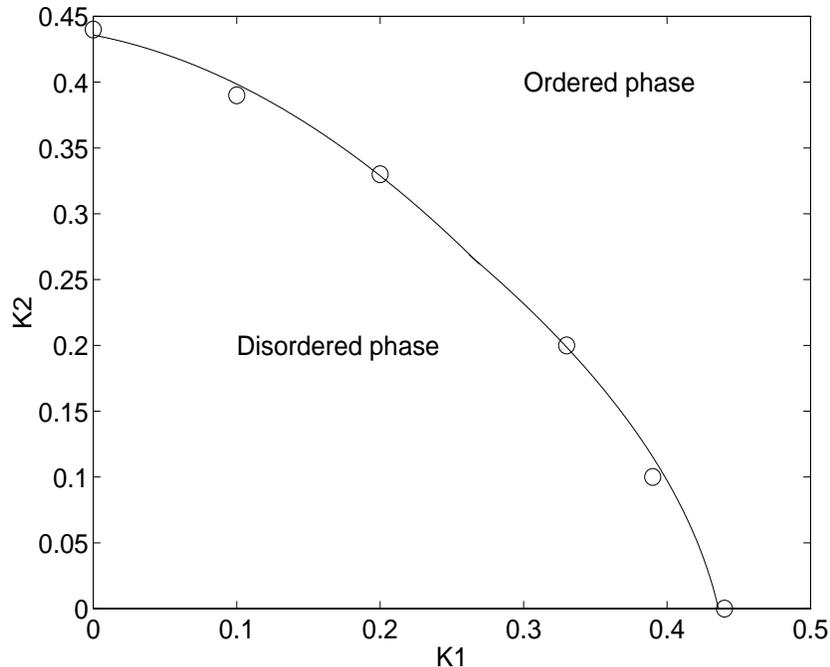}
$$
\caption{ \em Phase diagram for $ y=0.50 $. The lattice size is $32^2$.
         The solid line is an interpolation to the points.    } 

\label{fig:phase}
\end{figure}

First we need to determine the phase diagram of the model. For arbitrary
$\kappa_1$, $\kappa_2$ and $y$, Eq. (\ref{eq:model}) describes a two layered
Ising model with different couplings in each layer. In general, we expect that
the magnetization in one layer will induce the magnetization in the second
layer when the coupling between layers, $y$, is nonzero.  Therefore, the size
of the symmetric region,as shown in Figure~\ref{fig:phase},  will be reduced. A
few limiting cases can be easily derived.

If $y \ne 0, \kappa_2 = 0, \kappa_1 \ne 0$, the sum over $s_x$ can be
carried out explicitly
\begin{eqnarray}
Z &=& \sum_{\{\sigma\}} \sum_{\{s\}} e^{\textstyle \kappa_1 \sum_{x,\mu} 
\sigma_x\sigma_{x + \mu} + y \sum_x \sigma_x s_x} \\ \nonumber
&=& \sum_{\textstyle \{\sigma\}} \;\;  e^{ \textstyle \kappa_1 \sum_{x,\mu} \sigma_x\sigma_{x + \mu}}
e^{\sum_x ln [ ch (y\sigma_x) ] } \\ \nonumber
&=& const \times \sum_{\{\sigma\}} e^{\textstyle  \kappa_1\sum_{x,\mu}\sigma_x
\sigma_{x + \mu}} ~,\\
\end{eqnarray}
Thus the model becomes the standard 2d Ising model and the critical point
is known from the exact solution $\kappa_{1,cr} = \kappa_c = 0.44068...$.
It is also easy to show that a spontaneous symmetry breaking in $\sigma$
will induce a magnetization in $s$
\begin{equation}
<s> = tanh(y) <\sigma> ~,
\end{equation}
where $<\sigma>$ is given by the 2d Ising model solution.
Similar result can be derived for $\kappa_1 = 0, \kappa_2 \ne 0, y \ne 0$.

In the limit $y \to \infty$, $\sigma_x$ would be completely aligned with
$s_x$, and up to a constant the action can be written as
\begin{equation}
S = -(\kappa_1 + \kappa_2) \sum_{x,\mu} \sigma_x\sigma_{x + \mu} ~.
\end{equation}
This is again a 2d Ising model with critical point given by a straight line,
\begin{equation}
\kappa_1 + \kappa_2 = \kappa_c = 0.44068...~,
\end{equation}
in the $\kappa_1-\kappa_2$ phase plane.
At finite $y$, the phase boundary lies between the limit of $y = 0$ and
$y \to \infty$ as shown in Figure~\ref{fig:phase}. It was determined using Monte Carlo
simulations which were carried out on a 64-node CM-5 at Boston university.
For simplicity, we have used the heat-bath algorithm. The size of lattices
we used range from $16^2$ to $128^2$. The phase boundary was determined by 
observing the magnetization, defined as:
\begin{equation}
<m_1> = <\left|{1 \over L^2}\sum_x \sigma_x\right|>, 
\ \ \ \ <m_2> = <\left|{1 \over L^2}\sum_x s_x\right|>~.
\end{equation}
It is non-vanishing on a finite lattice in both phases, but will converge
to the physical magnetization in the infinite volume limit. Near the phase
transition point, $<m_1>, <m_2>$ changes rapidly from finite values to
almost zero in the symmetric phase. This cross-over behavior becomes
sharper with increasing lattice size. 
The results for  $y=0.50$ are shown in Figure~\ref{fig:phase}.
The points have been computed while the line is a simple interpolation. 
>From this figure one can get that $y=0.50$, $\kappa_2 = 0.20$ and
$\kappa_1 = 0.325(5)$ is a critical point. This point will be used in
the study of the propagators.
For a more precise measurement of the critical point, we used the Binder
index \cite{Binder}

\begin{equation}
U_1(L) = 1 - {<m_1^4> \over 3 <m_1^2>^2}~, \ \ \ \
U_2(L) = 1 - {<m_2^4> \over 3 <m_2^2>^2}~.
\end{equation}

\begin{figure}
\epsfxsize=11.0cm
\epsfysize=9.0cm
$$
\epsfbox{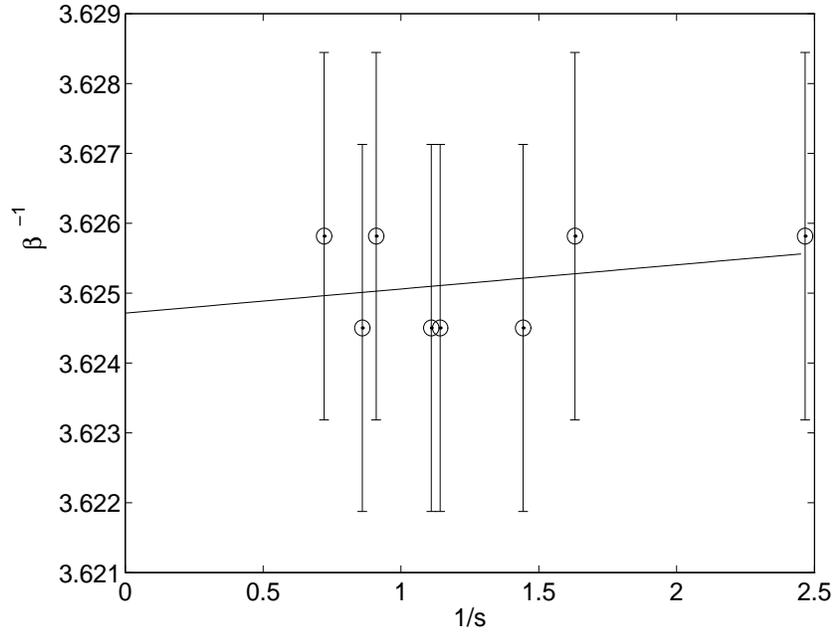}
$$
\caption{ \em Determination of the critical point using
          the finite size scaling of the
          Binder Index. The fitting curve is $ 1/\beta = a/s + 1/\beta_c $
          ($s=log(L1/L2)$).
          The computed values of the fitting parameters are:
          $\beta_c = 0.2759(2)$ and $a= 0.000(2)$.   } 

\label{fig:binder}
\end{figure}

\begin{figure}
\epsfxsize=11.0cm
\epsfysize=9.0cm
$$
\epsfbox{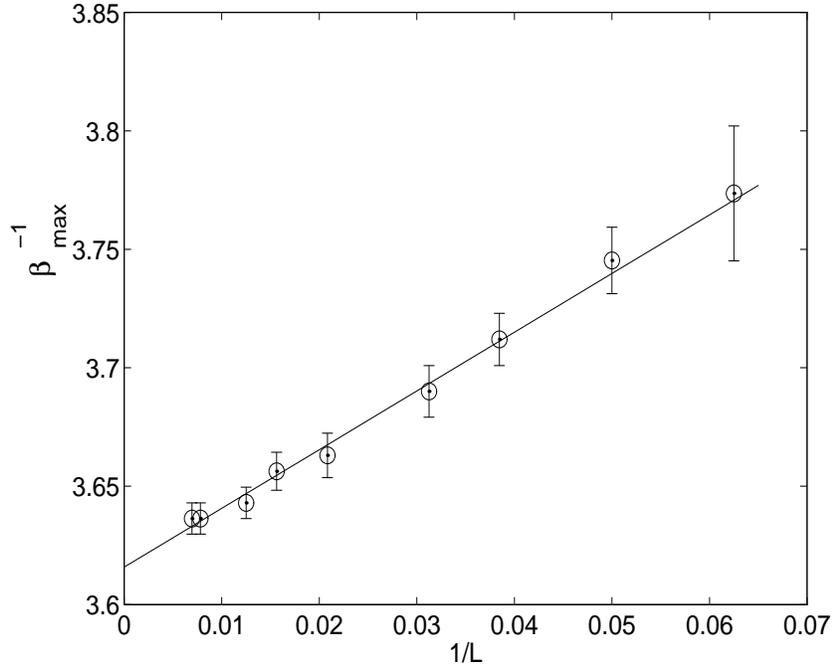}
$$
\caption{\em Location ($\beta_{max}$) of the Susceptibility Maximum vs.
          $L^{-\nu}$. Using Eq.\protect{\ref{eq:sumax}} 
          ($1/\beta_{max} = 1/\beta_c + a/L^{-\nu} $ ) we compute the
          critical point $\beta_c = 0.2766(4)$. 
          ($\nu$ is assumed to be $\nu=1$)
           } 
\label{fig:sumax}
\end{figure}

For the simplest case $\kappa_1 = \kappa_2 = y/2 = \beta$, we have
$U_1 = U_2 \equiv U(L,\kappa)$. When the volume is large, $U(L, \kappa)$
should cross at an unique point $(U^*, \kappa_c)$. However, for practical
$L$ and $L^\prime$, the crossing point depends on the ratio
$s = L^\prime / L$ due to residual finite size corrections. We obtained
the crossing point for a series of values of $s$, and fitted to the
form \cite{Binder}
\begin{equation}
{1 \over \beta} = {a \over ln(s)} + {1\over \beta_c}~,
\label{eq:binder}
\end{equation}
as shown in Figure~\ref{fig:binder}. We found $\beta_c = 0.2759(2)$.
Another way to locate the critical point is to use the finite size scaling
behavior of the susceptibility \cite{FSS}~\cite{Holm}.
We fitted the location of the peak ($\beta_{max}$) of susceptibility to:
\begin{equation}
{1 \over \beta_{max} } = {1 \over \beta_c } + {c \over L^{1\over \nu}}~,
\label{eq:sumax}
\end{equation}
with $\nu = 1$ (see next section). We found (see Figure~\ref{fig:sumax})
$\beta_c = 0.2766(4)$, which is consistent with the result obtained  
using the Binder index.
To be conservative, we take the difference in the above two estimates as an
indication of the true error and quote the critical point to be $\beta_c =
0.2759(7)$. 

As an aside, we would like to point out that our model can also be used to
test a very interesting method for computing critical points that has be
suggested recently~\cite{Wosiek}.  This method is based on the assumption that
the critical point of a system in $d$ dimensions is located at the maximum of
\begin{equation}
\rho(\beta) = \lim_{L\rightarrow\infty} \left( {(Tr {\cal T})^2 \over
Tr {\cal T}^2 } 
\right )^{\textstyle L^{1-d}} ,  \label{eq:wosiek}
\end{equation}
where $\cal T$ and ${\cal T}^2$ are the transfer matrices for the $d~-~1$
dimensional system and two coupled $d~-~1$ dimensional system respectively.
It has been shown that the above method is valid for self-dual models.
However, for non self-dual models it remains a conjecture.  Unfortunately, we
found from our simulations that the critical point of the two layered Ising
model is $\beta_c = 0.2759(7)$ so that it does not appear to agree with the
critical point obtained in Ref~\cite{Wosiek}.

\section{Critical Indices}

To establish the universality class, we need to investigate the finite
size scaling (FSS)~\cite{Holm}\cite{FSS} behavior and determine the critical indices.

\begin{figure}
\epsfxsize=11.0cm
\epsfysize=9.0cm
$$
\epsfbox{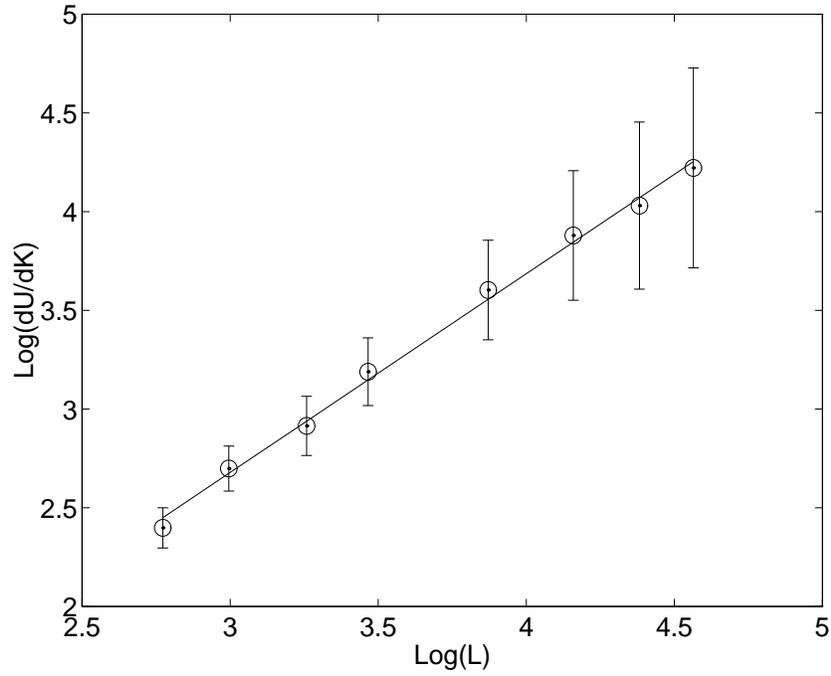}
$$
\caption{ \em From the finite size scaling  of 
           ${dU_L \over d \beta } \sim L^{1/\nu}$
            one obtains the critical exponent 
           $\nu = 0.99(3)$.}

\label{fig:dudk}
\end{figure}

\begin{figure}
\epsfxsize=11.0cm
\epsfysize=9.0cm
$$
\epsfbox{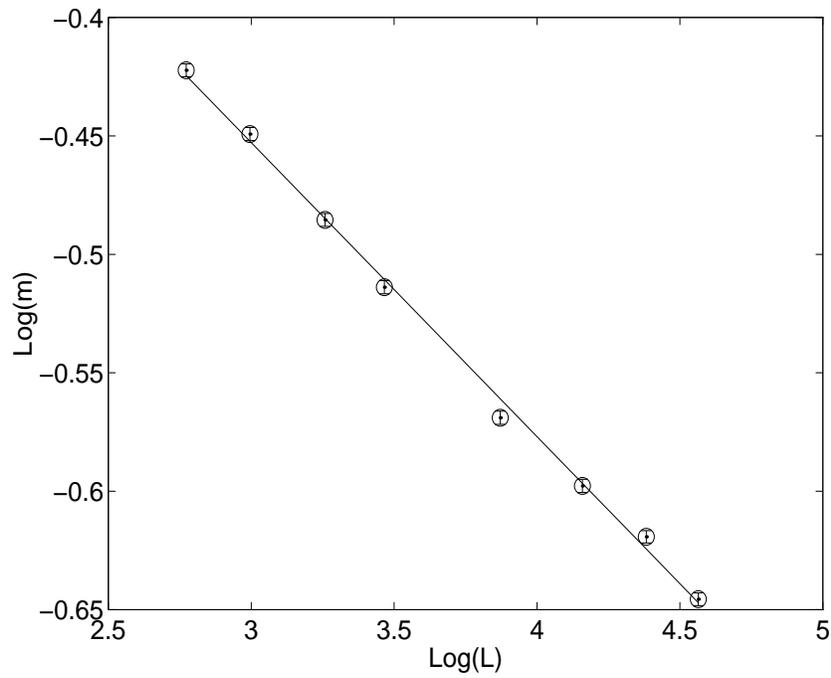}
$$
\caption{\em From the finite size scaling  of 
          $ M \sim L^{\beta/\nu}$ one obtains   
          $\beta / \nu = 0.124(2)$.}

\label{fig:logM}
\end{figure}

\begin{figure}
\epsfxsize=11.0cm
\epsfysize=9.0cm
$$
\epsfbox{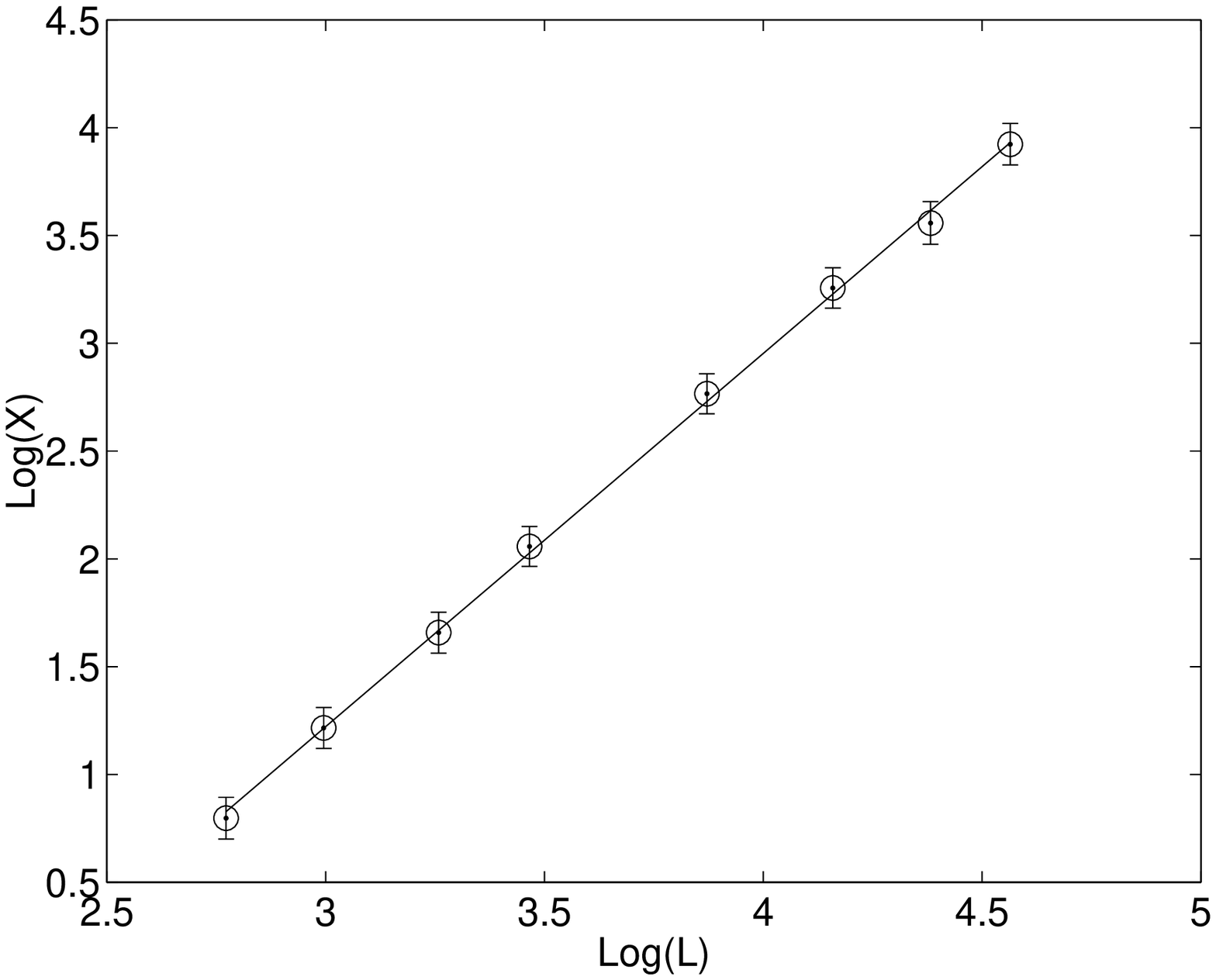}
$$
\caption{\em From the finite size scaling  of 
          $ X \sim L^{2-\eta}$
           one obtains  the anomalous dimension
          $\eta = 0.27(6)$.}

\label{fig:logX}
\end{figure}

For $\kappa_1 = \kappa_2 = y/2 = \beta$, we determine $\nu$ using
\begin{equation}
{dU_L \over d\beta} = (1- U_L)\left(<E> - 2 {<m^2 E> \over <m^2>}
+ {<m^4 E> \over <m^4>}\right)~,
\label{eq:dudk}
\end{equation}
where $E$ is the Ising energy per site. At $\beta_c$, $dU_L/d\beta$
should scale as $L^{1/\nu}$. As shown in Figure~\ref{fig:dudk}, if we take
$\beta_c = 0.2759$, we find $\nu = 0.99(3)$. This is to be compared
with the exact value $\nu = 1$ for the 2d Ising model.
Next we measured the magnetization at the critical point $\beta_c = 0.2759$.
Fitting to the scaling form
\begin{equation}
m \sim L^{-\beta /\nu}~,
\end{equation}
as shown in Figure~\ref{fig:logM}, we get $\beta /\nu = 0.124(2)$. This should be compared
to the exact value of 0.125
To get the anomalous  dimension, $\eta$, we measured the value of the
susceptibility at $\beta_c = 0.2759$ as a function of $L$ and fitted to
the form
\begin{equation}
\chi_{max} \sim L^{2 - \eta}~.
\end{equation}
As shown in Figure~\ref{fig:logX}, we found $\eta = 0.27(6)$ in 
comparison to the exact value of 0.25.
 All the results on the critical exponents are summarized in 
Table~\ref{tab:CrEx}. The rest of the critical indices
can be computed through scaling laws~\cite{Holm} 
$\alpha = 2 - \nu d$ , $\gamma / \nu = 2 - \eta $ ,
$\delta = (d+2-\eta) / (d-2+\eta)$.
Thus we conclude that all the critical indices of 
our model are the same as those
of the 2d Ising model. Consequently these two models are in the same 
Universality class and they have to the same continuum limit: a 
2d free Fermion system~\cite{Zuber}.

\begin{table}

\begin{center}
\begin{tabular}{||c||c|c||}\hline

Critical Exp.      &  2d Ising Model    & 2d Coupled Ising Model  \\ \hline
$ \nu $            &  1                 &  0.99(3)                \\ \hline
$\beta \over \nu$  &  0.125             &  0.124(2)               \\ \hline
$\eta$             &  0.25              &  0.27(6)                \\ \hline
$\alpha $          &  0                 &  0.02(6)                \\ \hline 
$\delta  $         &  15                &  14(3)                  \\ \hline
$\gamma \over \nu$ &  1.75              &  1.73(6)                \\ \hline
\end{tabular}
\vspace{0.6cm}
\label{tab:CrEx}
\end{center}
\caption{\em The critical exponents of the 2d Ising model and
 the coupled 2d Ising model. They seem to be identical with in the accuracy of
 our computation.}

\end{table}

\section{Mass Spectrum}

 We have established that the coupled Ising model in
Eq. (\ref{eq:model}) and the 2d Ising model obey the same finite
size scaling laws and therefore belong to the same universality class.
However the mass spectrum still remains a puzzle. Although there are
two fields $\sigma_x$ and $s_x$, we would like to know if there is one 
or two particles in the low energy spectrum.

\begin{figure}
\epsfxsize=11.0cm
\epsfysize=9.0cm
$$
\epsfbox{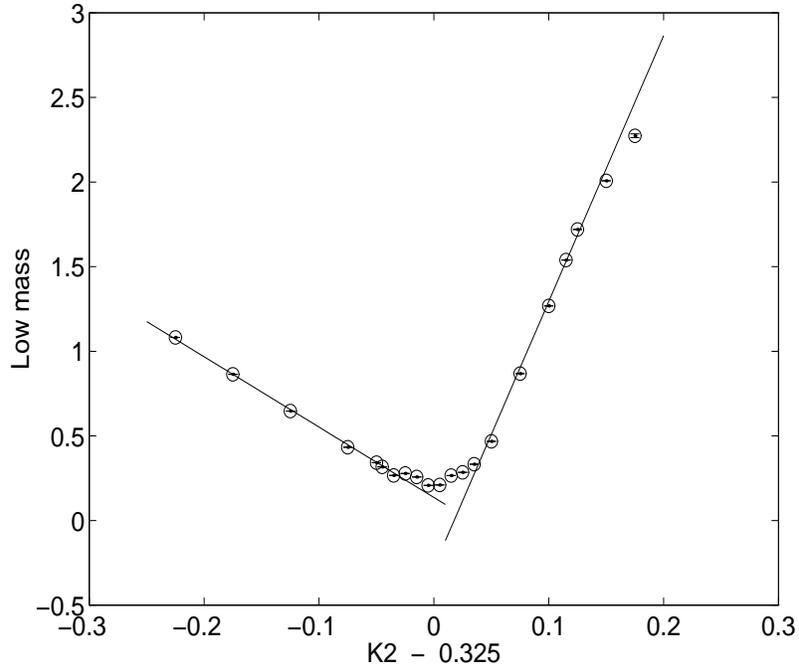}
$$
\caption{\em The linear fitting of the low mass around the critical point
         confirms the fact that $\nu$ is $1$. The mass unit is 
          $1/a$. }
\label{fig:sqmass2}
\end{figure}

Let us introduce the Fourier transformation of the fields as
\begin{equation}
\sigma(p) = {1 \over L^{d/2}} \sum_x \sigma_x e^{-ipx}~,
\end{equation}
\begin{equation}
s(p) = {1 \over L^{d/2}} \sum_x s_x e^{-ipx}~.
\end{equation}
The propagators in Fourier space can be written as
\begin{equation}
G(p) = \left( \begin{array}{cc}
<\sigma(p)\sigma(-p)> & <\sigma(p)s(-p)> \\
<s(p)\sigma(-p)>      & <s(p) s(-p)>     \\
\end{array} \right)~.
\end{equation}
Near $p^2 = 0$, the two-point function can be expanded as
\begin{equation}
G^{-1}(p) = A + B p^2 + O(p^4)~,
\label{eq:prop}
\end{equation}
where $A$ and $B$ are two-by-two matrices. $B$ leads to  a 
wavefunction renormalization. Thus the matrix $M=B^{-1}A$ 
is the mass matrix. The eigenvalues of $M$ are the masses of the
model (Here for simplicity, we define the masses at the
zero momentum instead of the pole of the propagator).
\begin{figure}
\epsfxsize=11.0cm
\epsfysize=9.0cm
$$
\epsfbox{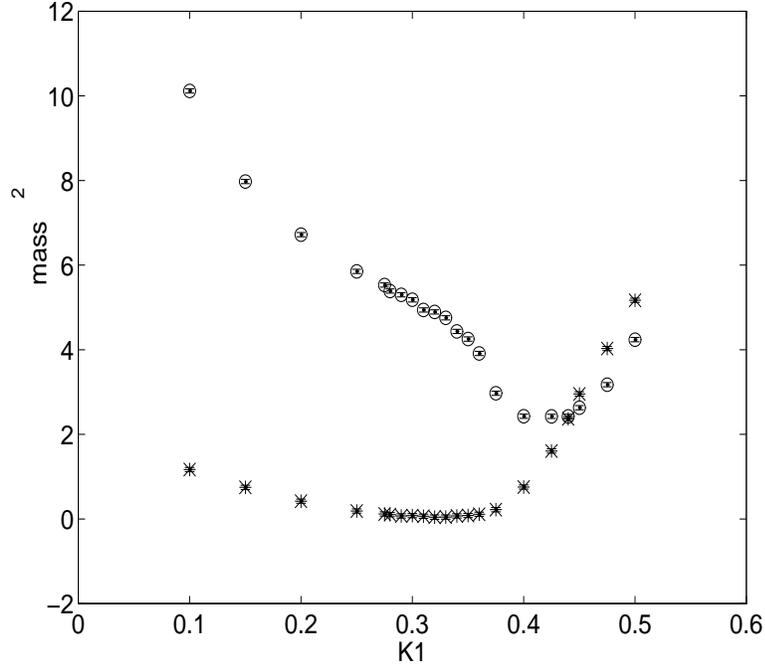}
$$
\caption{\em The mass matrix eigenvalues at $\kappa_2 = 0.20$ and $y=0.50$. 
         One goes to zero while the other remains 
         finite at the critical point $\kappa_1 = 0.325$. The mass unit is 
          $1/a$. }
\label{fig:mass}
\end{figure}
 Numerically, we have found that it is fairly easy to fit
$G^{-1}_{\sigma\sigma}(p)$, $G^{-1}_{ss}(p)$ 
near $p^2 = 0$ to a linear form, and $G^{-1}_{\sigma s}(p)$ is essentially
independent of $p^2$. We have obtained matrices $A$ and $B$
in Eq. (\ref{eq:prop}) for $y=0.50$ , $\kappa_2 = 0.20$ and for
various values of  $\kappa_1$ so that we cross the phase boundary.
The results for mass eigenvalues are shown in Figure~\ref{fig:mass}.
 It is very clear that while the small mass eigenvalue goes to
zero at the critical point, the heavy mass stays at $O(1)$ in lattice unit.
We also fitted the light mass eigenvalue to the scaling form
\begin{equation}
m^2 \sim (\kappa - \kappa_c)^{1/\nu}~,
\label{eq:mass}
\end{equation}
and found that for $\nu = 1$ one gets a very good fit of the data.
(see Figure~\ref{fig:sqmass2}.) The deviation form the linear fit close
to the critical point is due to FSS effects.
This is another indication that the critical exponent $\nu$ is indeed $1$
as for the 2d Ising model.

From the analysis of the propagator, we can learn two important things:
(1) only one particle remains in the physical spectrum even though
there appears to be extra degrees of freedom at the cut-off scale;
(2) the physical field is a linear combination of  bare fields
$\sigma_x$ and $s_x$.

\section{Conclusions}

In this paper we have studied a 2d spin model composed of two Ising systems
interacting through a Yukawa coupling.  There were two main objectives of our
effort.  One was to show that this form of coupling leads to the same
universality class as a single 2d Ising model; the other was to show that only
a single set of spin waves survives in the continuum limit.  Our study of the
critical exponents does support universality.  All critical exponents agree to
the accuracy of our computation.  In addition the analysis of the two point
correlation function supports the assertion that only one mass that goes to
zero at criticality or that the continuum theory has a single mass scale.  It
is encouraging that this can be observed numerically in a two channel
correlations function, when in reality the scalar excitations do not lie in a
two dimensional subspace, but must include the multibody states. Nonetheless
the freezing out of extra degrees of freedom is observable.  Thus we were able
to get solid numerical evidence that the continuum theory for the two spin
Yukawa model is identical to the continuum limit of a single 2d Ising model.

This result is not surprising from the viewpoint of renormalization group. One
can think the integration over one set of spins as a ``blocking''
transformation which will induce irrelevant operators for the action of the 2d
Ising model. As we noted in the introduction this result is almost
``obvious'', but we are reassured by numerical ``proof'' and encourage to
extend these numerical method to subtler coupled Higgs systems where our
plausibility arguments might not be convincing. Our model does serves as an
illustration for the key feature of Chirally Extended
QCD~\cite{BST}~\cite{BOT} which, although involves extra degrees of freedom on
the lattice, is expected to lead to the same continuum theory as the standard
Wilson Lattice QCD.  The numerical success of this demonstration gives us hope
that more complex coupled systems can be studied by similar techniques. In the
near future, we will attempt to extend these methods to 2d and 4d Yukawa
coupled chiral spin models.


\begin{thebibliography}{99}

\bibitem{Kogut} For example, see, J.~B.~Kogut, Rev. Mod. Phys. 51 (1979) 659.

\bibitem{Coleman} S.~Coleman, Phys. Rev. D 11 (1974) 2911.

\bibitem{UCSD}A.~Hasenfratz, P.~Hasenfratz, K.~Jansen, J.~Kuti
and Y.~Shen, Nucl. Phys. B365 (1991) 79.

\bibitem{BST} R.~C.~Brower, Y.~Shen and C.~I.~Tan, preprint BUHEP-94-3.

\bibitem{BOT} R.~C.~Brower, K.~Orginos and C.~I.~Tan, Nucl. Phys. B 
(Porc. Suppl.) 42(1995) 42.

\bibitem{Binder} K.~Binder, Z. Phys. B43 (1981) 119.
M.~S.~S.~Challa, D.~P.~Landau and K.~Binder, Phys. Rev. B34 (1986) 1841.

\bibitem{FSS} For a review on finite size scaling, see, M.~N.~Barber,
in {\it Phase Transitions and Critical Phenomena}, Ed. C.~Domb and
J.~L.~Lebowitz, Vol 8, p145. Academic, New York, 1983.

\bibitem{Wosiek} Z.~Burda and J.~Wosiek, Nucl. Phys B (Proc. Suppl.)
34 (1994) 667.


\bibitem{Holm} C.~Holm and W.~Janke Nucl. Phys. B (Proc. Suppl.) 30 
                  (1993) 846.

\bibitem{Zuber} J.~B.~Zuber, Phys.  Rev. D 15 (1977) 2875.




\end{thebibliography}
\end{document}